\documentclass[conference]{IEEEtran}
\IEEEoverridecommandlockouts
\usepackage{cite}
\usepackage{amsmath,amssymb,amsfonts}
\usepackage{algorithmic}
\usepackage{graphicx}
\usepackage{textcomp}
\usepackage{xcolor}
\usepackage{tabularx}
\usepackage{caption}
\usepackage{hyperref}
\usepackage{url}
\usepackage{cite}

\hypersetup{
    colorlinks=true,
    citecolor=cyan,
    linkcolor=blue,
    urlcolor=blue
}

\def\BibTeX{{\rm B\kern-.05em{\sc i\kern-.025em b}\kern-.08em
    T\kern-.1667em\lower.7ex\hbox{E}\kern-.125emX}}
\begin{document}

\title{EMelodyGen: Emotion-Conditioned Melody Generation in ABC Notation with the Musical Feature Template}

\author{Monan Zhou$^1$, Xiaobing Li$^1$, Feng Yu$^1$, Wei Li$^{2,3*}$\thanks{* Corresponding author.}\\\small
  $^1$Department of Music AI and Information Technology, Central Conservatory of Music, Beijing, China\\
  $^2$School of Computer Science and Technology, Fudan University, Shanghai, China\\
  $^3$Shanghai Key Laboratory of Intelligent Information Processing, Fudan University, Shanghai, China
}

\maketitle
\begin{abstract}
  The EMelodyGen system focuses on emotional melody generation in ABC notation controlled by the musical feature template. Owing to the scarcity of well-structured and emotionally labeled sheet music, we designed a template for controlling emotional melody generation by statistical correlations between musical features and emotion labels derived from small-scale emotional symbolic music datasets and music psychology conclusions. We then automatically annotated a large, well-structured sheet music collection with rough emotional labels by the template, converted them into ABC notation, and reduced label imbalance by data augmentation, resulting in a dataset named Rough4Q. Our system backbone pre-trained on Rough4Q can achieve up to 99\% music21 parsing rate and melodies generated by our template can lead to a 91\% alignment on emotional expressions in blind listening tests. Ablation studies further validated the effectiveness of the feature controls in the template. Available code and demos are at \href{https://github.com/monetjoe/EMelodyGen}{here}.
\end{abstract}


\section{Introduction} \label{sec:intro}
Recent emotion-conditioned music generation approaches such as \cite{ji2024muser,zheng2024real,huang2024emotion,sulun2022symbolic,ferreira2022controlling,grekow2021monophonic} use MIDI data instead of ABC notation for symbolic music generation, and some of them are not strictly studied in Russell 4Q \cite{russell1980circumplex} emotional label system, while our work attempts to generate ABC notation melodies in Russell 4Q emotional space controlled by the musical feature template, so there is no direct comparative task available. We chose to use ABC notation because of its higher musical information density compared to sheet MIDI and XML/MusicXML/MXL. In the field of sheet music generation in ABC notation, Tomasz Michal Oliwa previously explored genetic algorithms for rock music composition \cite{10.1145/1389095.1389399} instead of emotionally conditioned composition. More recently, approaches such as Tunesformer \cite{wu2023tunesformer}, abcMLM \cite{casini2024investigating}, and MelodyT5 \cite{wu2024melodyt5} have utilized transformer-based language models to generate music in ABC notation without emotional conditions. Among these, abcMLM and MelodyT5 are on the basis of transformer encoder, encoder-decoder architecture, respectively. So we selected Tunesformer, which is a transformer decoder-only model, as the backbone of our system because of its focus on generation. However, the effectiveness of Tunesformer in generating sheet music heavily relies on the quality of the training data, so models trained by disorganized sheet music may generate erroneous sheet music that cannot be correctly parsed by music21\footnote{\href{https://pypi.org/project/music21}{https://pypi.org/project/music21}} and then properly rendered into audios.

So far, well-structured sheet music data with emotional labels is scarce. Notable datasets such as EMOPIA \cite{hung2021emopia} and VGMIDI \cite{ferreira_ismir_2019} merely consist of MIDI data with emotional annotations instead of XML/ABC sheet music. Although performance MIDI files can be converted to XML/ABC formats by tools such as music21 or \textit{MuseScore}\footnote{\href{https://musescore.org}{https://musescore.org}}, the resulting scores are often disorganized, which will reduce the quality of melody generation. To address this issue, we selected well-structured sheet music as the foundational training material and subsequently exerted an emotional template for melody generation through the control of musical features. This approach not only ensures a certain level of quality in the generated output but also circumvents the issue of the absence of emotional labels in well-structured sheet music datasets. To identify key musical features that are highly correlated with emotional control, we merged the EMOPIA and VGMIDI datasets, both of which are annotated with emotional labels, into a unified analysis dataset. We then extracted features from it and did a correlation analysis between musical features and emotional labels. To mitigate potential biases that may arise from data distribution preferences, we also referenced prior knowledge from music psychology literature \cite{scherer2013music}, ultimately devising a musical feature template tailored for emotionally controlled melody generation. The selected features can be categorized into two classes: the controllable features that can be modified directly at the output stage, such as octave, volume, etc.; the embedded features that require deep learning for the model to interpret, such as pitch range, average pitch (avg pitch), pitch standard deviation (pitchSD), melodic ascending/descending (direction), mode, etc.

The designed template incorporates five features from the aforementioned categories, two of which belong to the embedded features that can merely achieve emotional control through embedding, so part of them need to be annotated into the training materials for fine-tuning. The Rough4Q dataset was just constructed through automatic annotation of the two features on the basis of a well-structured sheet music collection, and the two features were embedded into the backbone model via fine-tuning. The backbone pre-trained on Rough4Q achieved up to 99\% music21 parsing rate and the emotional alignment of the generated music by the template with human expectations reached 91\% in blind listening tests. Additionally, ablation experiments were conducted to validate the effectiveness of these five control conditions in the template on overall emotional expression. In summary, the principal contributions of this paper are listed as follows:
\begin{itemize}
  \item We conduct holistic correlation analysis between musical features and emotions to design the template for emotion control incorporating music psychology findings.
  \item We propose a 2-stage method to avoid label scarcity: auto-labeling well-structured datasets to gatekeep generation quality and using the template for emotion control.
  \item Our work is the first exploratory study for emotion-conditioned melody generation via ABC notation.
\end{itemize}

\section{Datasets} \label{sec:data}
Four datasets were created for various experiments: processed EMOPIA and processed VGMIDI were used to compare the music21 parsing rates of the backbone after being fine-tuned on these datasets versus well-structured sheet music; the analysis dataset was utilized to examine the correlations between various musical features and emotional labels; and Rough4Q was employed for embedding features to the final model via fine-tuning. The details of data structures and statistical summaries of these four datasets are described in subsequent subsections.

\subsection{Processed EMOPIA \& VGMIDI} \label{subsec:emopia}
\begin{figure}[htbp]
  \begin{minipage}[b]{0.49\linewidth}
    \centerline{\includegraphics[scale=0.36]{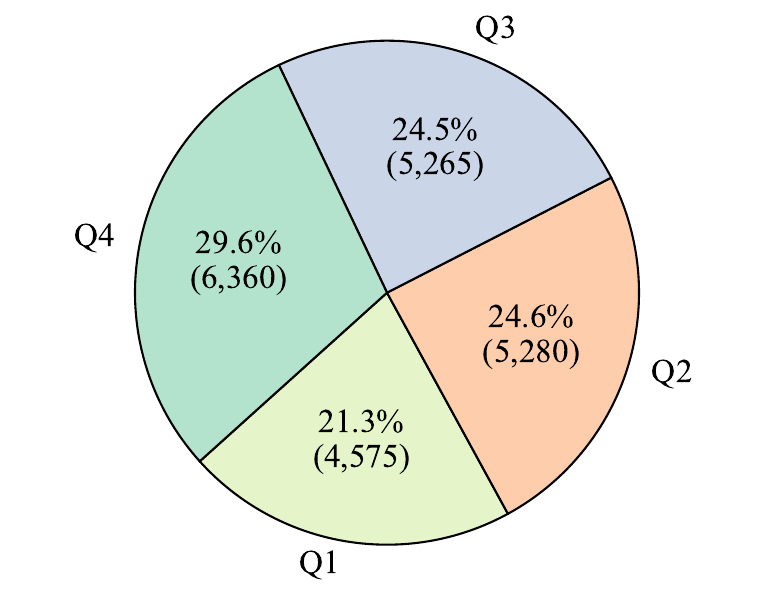}}
    \centerline{Processed EMOPIA}\medskip
  \end{minipage}
  \begin{minipage}[b]{0.49\linewidth}
    \centerline{\includegraphics[scale=0.36]{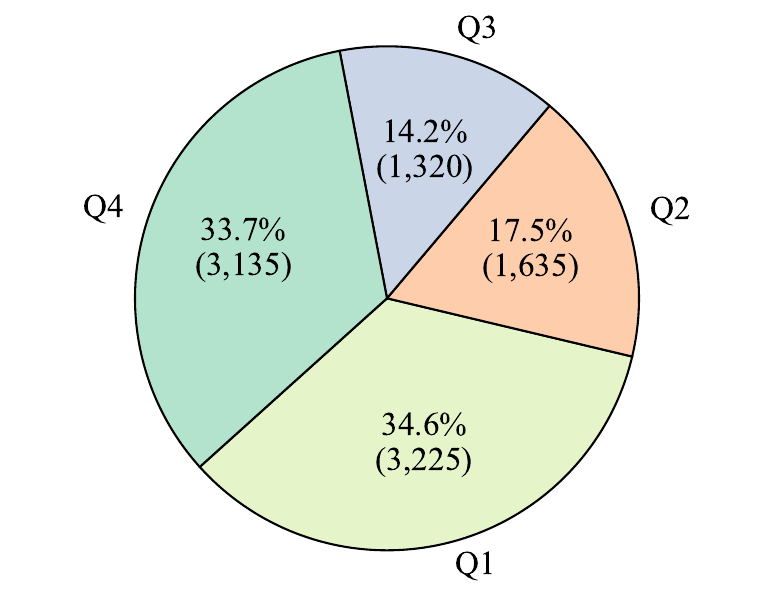}}
    \centerline{Processed VGMIDI}\medskip
  \end{minipage}\hfill
  \begin{minipage}[b]{0.49\linewidth}
    \centerline{\includegraphics[scale=0.36]{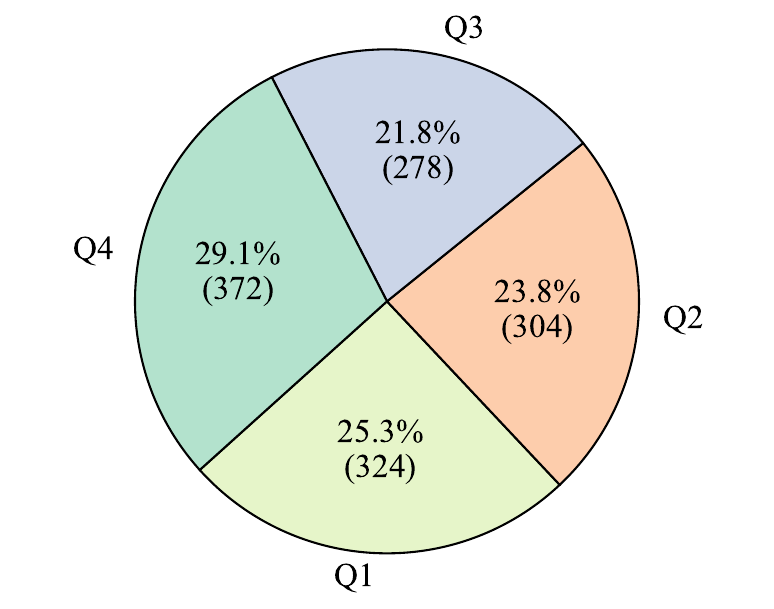}}
    \centerline{Analysis dataset}\medskip
  \end{minipage}\hfill
  \begin{minipage}[b]{0.49\linewidth}
    \centerline{\includegraphics[scale=0.36]{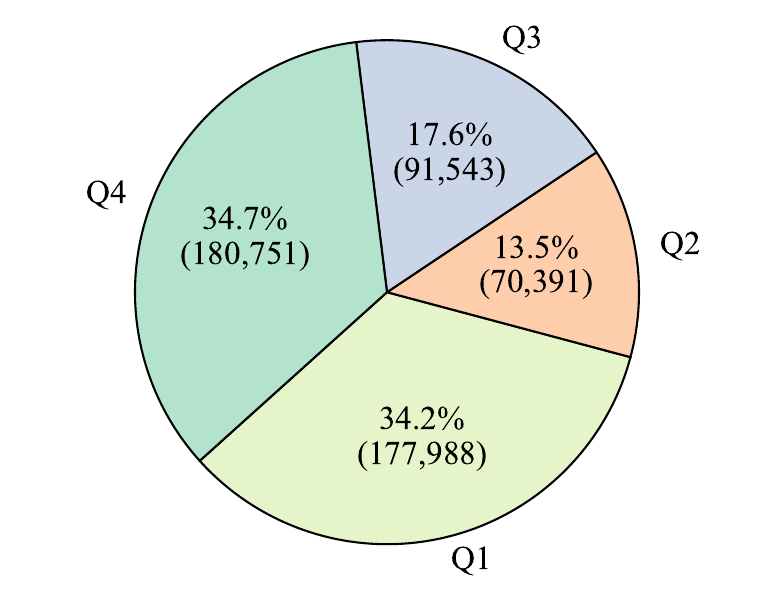}}
    \centerline{Rough4Q (proposed)}\medskip
  \end{minipage}
  \caption{Pie charts with proportions on different emotion categories of the processed datasets.}
  \label{fig:pies}
\end{figure}

It is essential to ensure that the two processed datasets are compatible with the input format required by the pre-trained backbone for their use. We found that the average number of measures in the dataset used for the pre-training backbone is approximately 20, and the maximum number of measures supported by the pre-trained backbone input is 32. Consequently, we converted the original EMOPIA and VGMIDI data into XML scores filtering out erroneous items, and segmented them into chunks of 20 measures. Each chunk was appended with an ending marker to prevent the model from generating endlessly in cases of repetitive melodies without seeing a terminating mark. For the ending segments of the scores, if a segment exceeded 10 measures, it was further divided; otherwise, it was combined with the previous segment. This approach ensures that the resulting score slices do not exceed 30 measures, thereby guaranteeing that all slices are within the maximum measure limit supported by the backbone, with an average of approximately 20 measures.
Subsequently, we converted the segmented XML slices into ABC notation format, performed data augmentation by transposing to 15 keys, and extracted the melodic lines with control codes to produce the final processed EMOPIA and VGMIDI datasets. Both datasets have a consistent structure comprising three columns: the first column is the Tunesformer's control code, the second column is ABC chars, and the third column contains the 4Q emotion labels inherited from the original dataset. The total number of data is 21,480 for processed EMOPIA and 9,315 for processed VGMIDI, which were both split into training and testing sets at a 10:1 ratio. Their label distributions are presented in the first to second subplots of Fig. \ref{fig:pies}.
\begin{table*}[htbp]
  \centering
  \caption{Pearson correlation statistics between emotions and features for the merged data from EMOPIA and VGMIDI.}
  \label{tab:pearson}
  \footnotesize
  \begin{tabularx}{\textwidth}{rl|cXXl}
    \hline
    \textbf{Emotion} & \textbf{Feature}                     & \textbf{Correlation coefficient} & \textbf{Relevance} & \textbf{P-value}   & \textbf{Confidence level}                \\\hline
    Valence          & Key                                  & +0.0123                          & Weak positive      & 6.594e-01          & $p \ge 0.05$ insignificant               \\
    \textbf{Valence} & \textbf{Mode}                        & \textbf{+0.3850}                 & \textbf{Positive}  & \textbf{2.018e-46} & \textbf{\textit{p} $<$ 0.05 significant} \\
    Valence          & Tempo                                & +0.0621                          & Weak positive      & 2.645e-02          & $p<0.05$ significant                     \\
    Valence          & Direction                            & +0.0010                          & Weak positive      & 9.709e-01          & $p \ge 0.05$ insignificant               \\
    Valence          & Avg pitch (guides octave control)    & +0.0102                          & Weak positive      & 7.161e-01          & $p \ge 0.05$ insignificant               \\
    Valence          & Pitch range                          & -0.0771                          & Weak negative      & 5.794e-03          & $p<0.05$ significant                     \\
    Valence          & PitchSD                              & -0.0676                          & Weak negative      & 1.568e-02          & $p<0.05$ significant                     \\
    Valence          & RMS (guides volume control)          & +0.1174                          & Weak positive      & 2.597e-05          & $p<0.05$ significant                     \\\hline
    Arousal          & Key                                  & -0.0007                          & Weak negative      & 9.809e-01          & $p \ge 0.05$ insignificant               \\
    Arousal          & Mode                                 & -0.0962                          & Weak negative      & 5.748e-04          & $p<0.05$ significant                     \\
    Arousal          & Tempo                                & +0.1579                          & Weak positive      & 1.382e-08          & $p<0.05$  significant                    \\
    Arousal          & Direction                            & -0.0958                          & Weak negative      & 6.013e-04          & $p<0.05$  significant                    \\
    Arousal          & Avg pitch (guides octave control)    & -0.1818                          & Weak negative      & 5.919e-11          & $p<0.05$  significant                    \\
    \textbf{Arousal} & \textbf{Pitch range}                 & \textbf{+0.3276}                 & \textbf{Positive}  & \textbf{2.324e-33} & \textbf{\textit{p} $<$ 0.05 significant} \\
    \textbf{Arousal} & \textbf{PitchSD}                     & \textbf{+0.3523}                 & \textbf{Positive}  & \textbf{1.179e-38} & \textbf{\textit{p} $<$ 0.05 significant} \\
    \textbf{Arousal} & \textbf{RMS (guides volume control)} & \textbf{+0.3800}                 & \textbf{Positive}  & \textbf{3.558e-45} & \textbf{\textit{p} $<$ 0.05 significant} \\\hline
  \end{tabularx}
\end{table*}

\subsection{The Analysis Dataset} \label{subsec:analysis}
The analysis dataset is derived from merging the original EMOPIA and VGMIDI into the Russell 4Q label system. This dataset consists of 11 columns: the first three columns are emotion labels, specifically label (Russell 4Q emotions), valence (low=0 or high=1), and arousal (low=0 or high=1); the remaining eight columns represent features, which are key (one of the 12 keys: \textit{C, C\#, D, Eb, E, F, F\#, G, G\#/Ab, A, Bb, B}), mode (minor=0 or major=1), direction (descending=0 or ascending=1), avg pitch (octave), pitch range, pitchSD, tempo, and RMS (volume).

For feature extraction, the key, mode, direction, avg pitch, pitch range, and pitchSD features were directly extracted by music21 at the symbolic level. However, tempo extracted from MIDI files often defaults to 120 BPM, which may not reflect the actual value, we used \textit{MuseScore} to render these MIDI files into WAV format with a 44.1 kHz sample rate by its default piano soundfont. Subsequently, we used the \textit{librosa}\footnote{\href{https://pypi.org/project/librosa}{https://pypi.org/project/librosa}} library to estimate the tempo as more accurate and distinguishable data, which also calculated the RMS of the rendered audios.

It is noted that the extraction of the latter two features (tempo and RMS) from rendered audio is less efficient compared to the first six features. However, the rendering time for these features is acceptable for the analysis phase because the combined dataset comprises only 1,278 pieces of music. Consequently, we have constructed the analysis dataset, with its distribution according to the Russell 4Q classification shown in the third pie chart of Fig. \ref{fig:pies}. For the statistical correlation analysis, we computed the correlations between the second and third columns (valence and arousal) and the remaining eight columns (features) and obtained Table \ref{tab:pearson}.

\subsection{Rough4Q Dataset} \label{subsec:4q}
This large-scale dataset was created by automatically annotating a substantial amount of well-structured sheet music on the basis of conclusions in Table \ref{tab:pearson} and the music psychology literature. The data sources for this dataset, detailed in Table \ref{tab:rough4q}, include both scores in XML/MXL/MusicXML and ABC notation format. After filtering out erroneous and duplicated scores, and consolidating them into a unified XML format, we rapidly extracted the two computationally manageable features, pitchSD and mode, by music21.
\begin{table}[htbp]
  \centering
  \footnotesize
  \caption{Comparison of source datasets for Rough4Q by size in ascending order.}
  \label{tab:rough4q}
  \begin{tabular}{lrc}
    \hline
    \textbf{Source dataset}                                              & \textbf{Size} & \textbf{Original format} \\\hline
    \textit{Midi-Wav Bi-directional Pop} \cite{zhaorui_liu_2021_5676893} & 111           & MusicXML                 \\
    \textit{JSBach Chorales} \cite{wu2023chord}                          & 366           & MXL                      \\
    \textit{Nottingham} \cite{nottingham}                                & 1,015         & ABC notation             \\
    \textit{Wikifonia} \cite{simonetta_2018_1476555}                     & 6,394         & MXL                      \\
    \textit{Essen Folk Song} \cite{esac}                                 & 10,369        & ABC notation             \\
    \textit{IrishMAN} \cite{wu2023tunesformer}                           & 216,281       & XML                      \\\hline
  \end{tabular}
\end{table}

To unify the labeling system between Rough4Q and other datasets, mode, and pitchSD were used as approximations for valence and arousal, respectively, owing to their strong positive correlations with these emotional dimensions. This also resulted in a 4Q label system, which is the reason for calling it Rough4Q too. The original dataset exhibited a severe class imbalance issue, with the data volume of the Q2 and Q3 categories being an order of magnitude lower than the other categories. To address this, 15-key data augmentation was applied specifically to the Q2 and Q3 categories. From the statistical conclusions in Table \ref{tab:pearson}, there is almost no correlation between key and emotion, indicating that transposing to 15 keys is unlikely to significantly impact the label distribution. The statistics of the final augmented dataset are presented in the last subplot of Fig. \ref{fig:pies}.

\section{Methodology} \label{sec:method}
In the correlation analysis phase, the employed methods included the Pearson correlation coefficient (PCC) \cite{pearson1985ii} and Gaussian kernel density estimation (KDE) \cite{parzen1962estimation}. The former was used to calculate the correlation coefficients between features and emotions, while the latter was utilized for plotting the distribution of emotion labels across features. The subsequent fine-tuning stage discusses details of the backbone network.

\subsection{Correlation Analysis} \label{subsec:pearson}
We performed statistical analysis on the analysis dataset in the Russell 4Q emotion label system, calculating the PCCs between emotional dimensions and features of sheet music with Gaussian KDE plots for multiscale features and bar plots for bisector features. Since EMOPIA and VGMIDI use 4Q and binary valence/arousal (V/A) labels, respectively. Both of them are not continuous values and can be converted between each other as follows:
\begin{equation}
  \mathcal{Q}(V, A)=I_{V<0} I_{A\ge 0} + 2 I_{V<0} I_{A<0} + 3 I_{V \ge 0} I_{A<0}
  \label{eq:4q}
\end{equation}
where $I$ is the indicator function. Therefore, we chose the negative and positive signs of V/A values as two levels: low and high, with values of 0 and 1, respectively.

Regarding the selection of musical features, we extracted eight musical features: key, mode, tempo, direction, avg pitch, pitch range, pitchSD, and RMS from the melodies. Among these, calculating avg pitch, pitchSD, and direction requires further explanation: assuming a melody $M = \{(p_1, d_1), (p_2, d_2), \ldots, (p_n, d_n)\}$ consists of $n$ notes, where $p_i$ and $d_i$ represent the pitch and duration of the $i$-th note, respectively. The durations here are extracted by music21, where 1 represents the length of a quarter note. The avg pitch (denoted as $\bar{p}$) is the weighted average of pitch by duration:
\begin{equation}
  \bar{p} = \frac{\sum_{i=1}^{n} p_i d_i}{\sum_{j=1}^{n} d_j}
  \label{eq:avg}
\end{equation}

On the basis of $\bar{p}$, we further calculated the pitchSD as the weighted standard deviation of pitches by duration:
\begin{equation}
  pitchSD = \sqrt{\frac{\sum_{i=1}^{n} (p_i - \bar{p})^2 d_i}{\sum_{j=1}^{n} d_j}}
  \label{eq:std}
\end{equation}

For the feature direction, since it describes a musical characteristic at the phrase level instead of entire pieces, we approach it by analyzing the duration of ascending and descending segments statistically. By comparing these durations, we determine the overall tonal direction of the piece. Specifically, if the total duration of ascending segments is greater, the piece is labeled as having an ascending tonal direction. Otherwise, the piece is labeled as having a descending tonal direction.

\begin{figure*}[htbp]
  \begin{minipage}[b]{0.24\linewidth}
    \centerline{\includegraphics[scale=0.2]{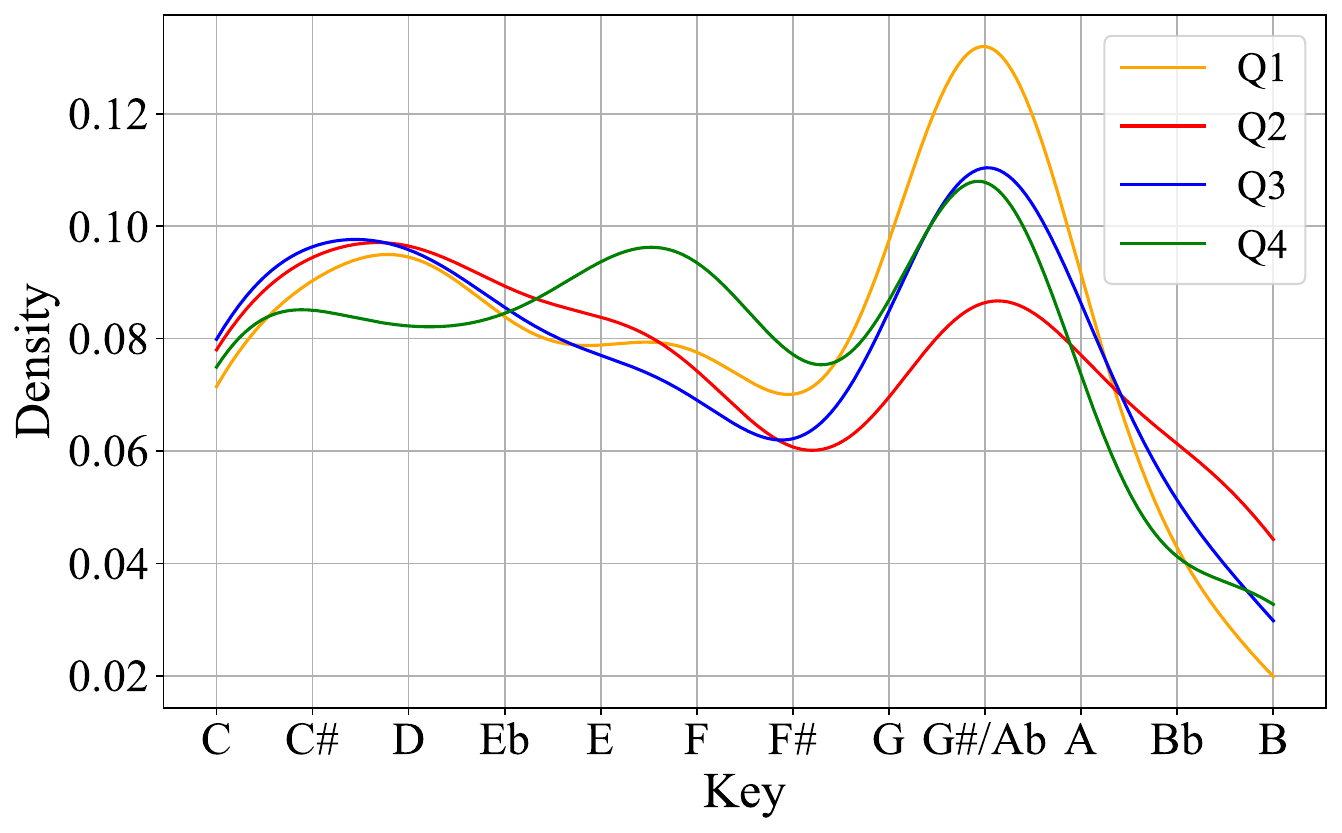}}\medskip
  \end{minipage}
  \begin{minipage}[b]{0.24\linewidth}
    \centerline{\includegraphics[scale=0.2]{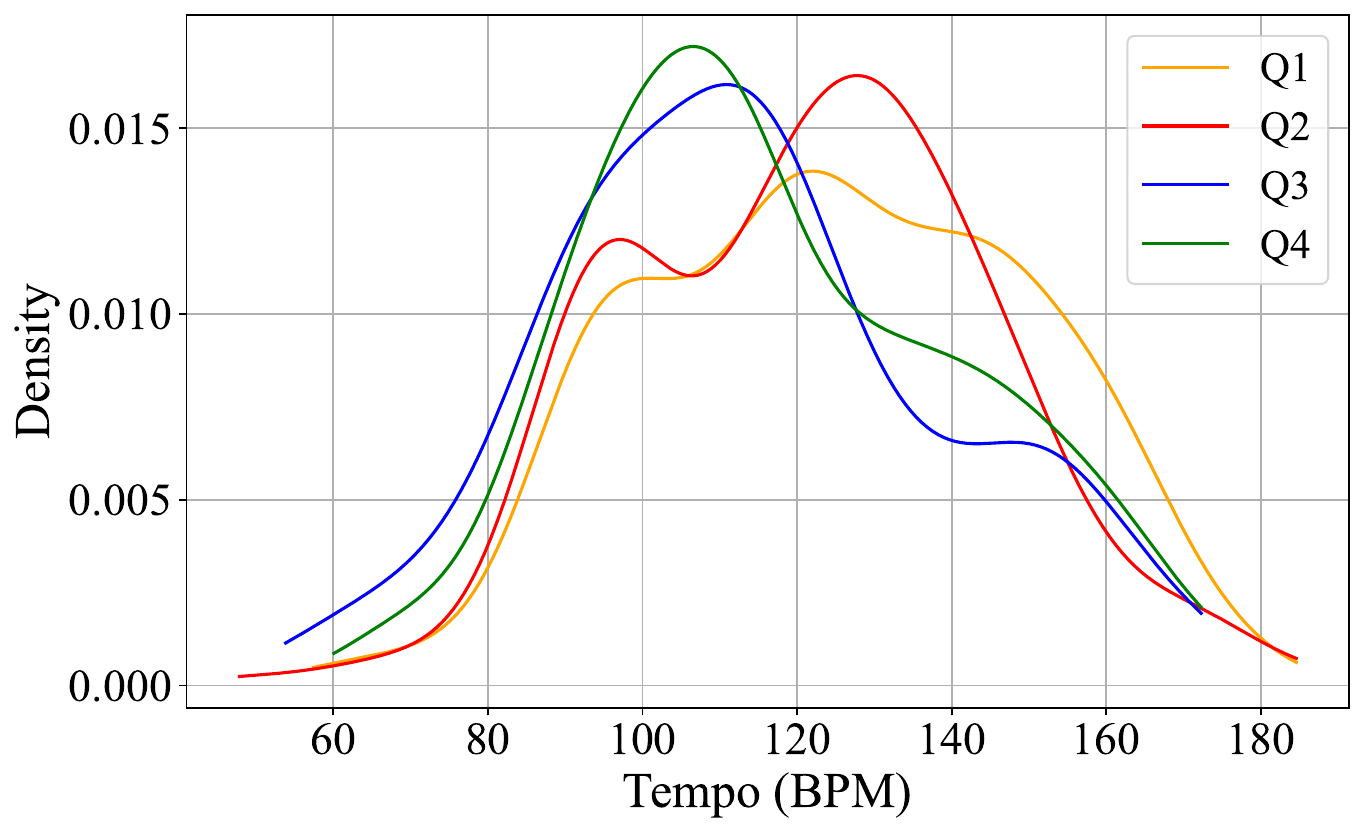}}\medskip
  \end{minipage}
  \begin{minipage}[b]{0.24\linewidth}
    \centerline{\includegraphics[scale=0.2]{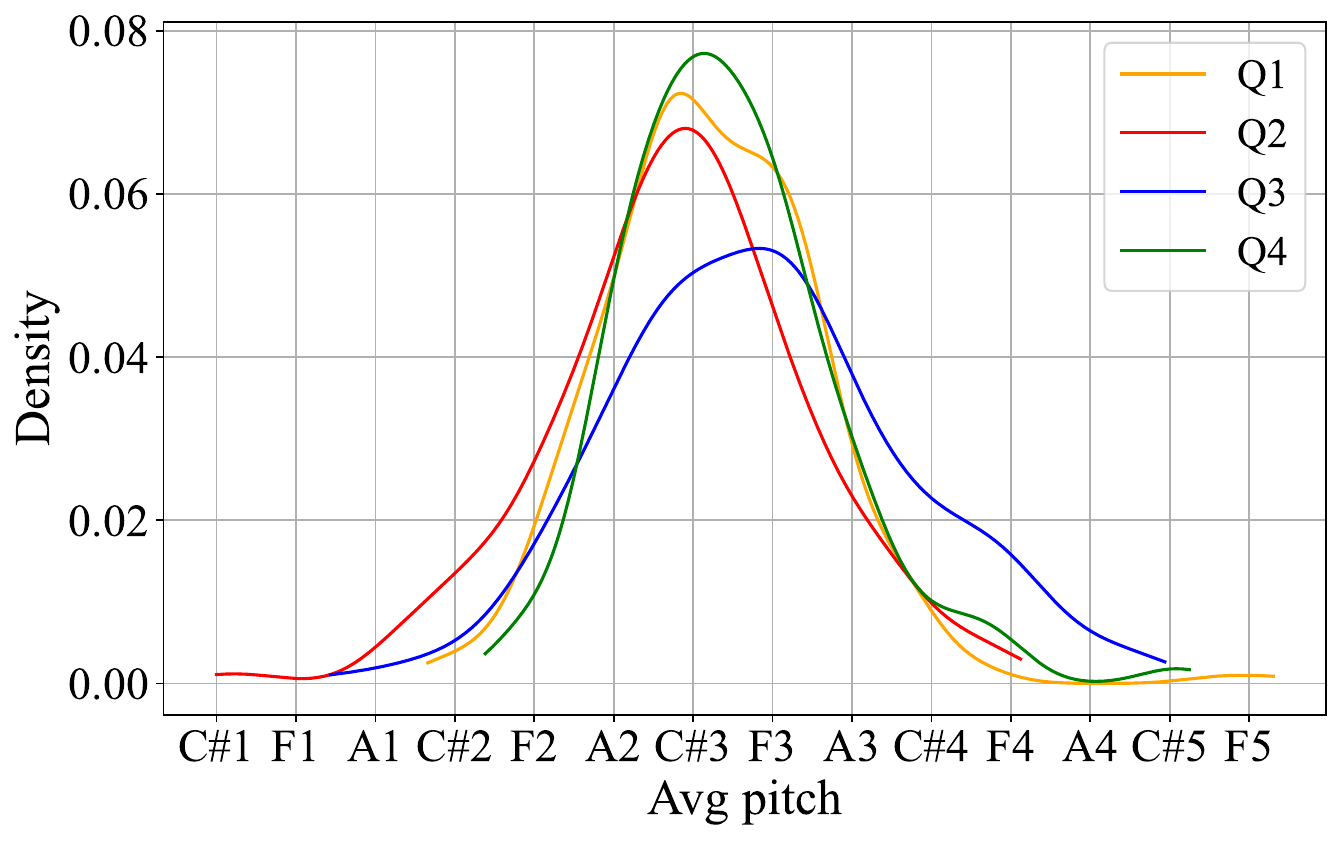}}\medskip
  \end{minipage}
  \begin{minipage}[b]{0.24\linewidth}
    \centerline{\includegraphics[scale=0.24]{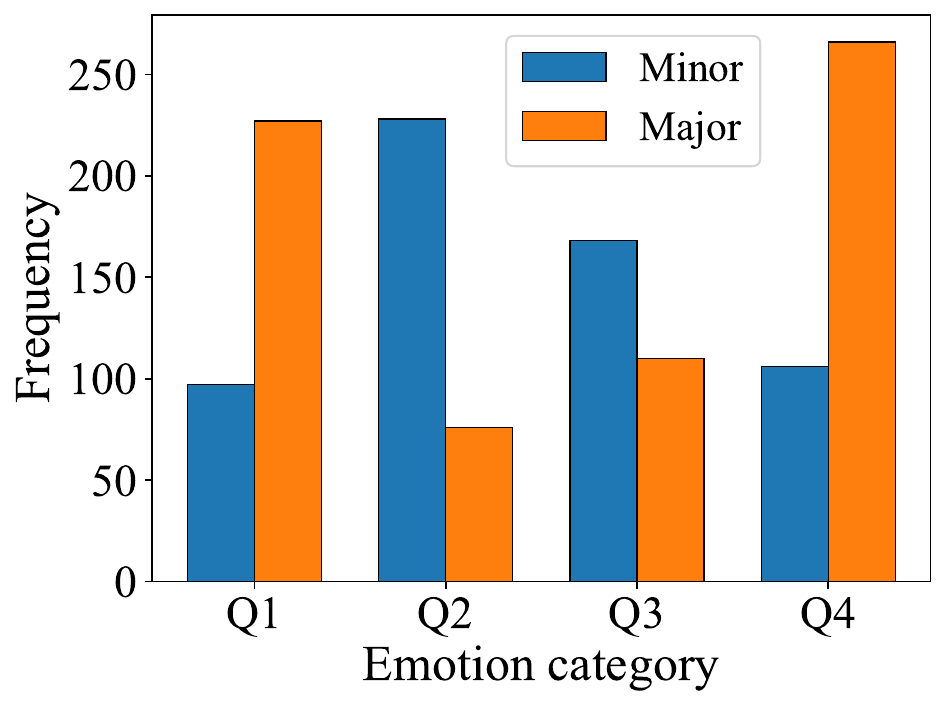}}\medskip
  \end{minipage}\hfill
  \begin{minipage}[b]{0.24\linewidth}
    \centerline{\includegraphics[scale=0.2]{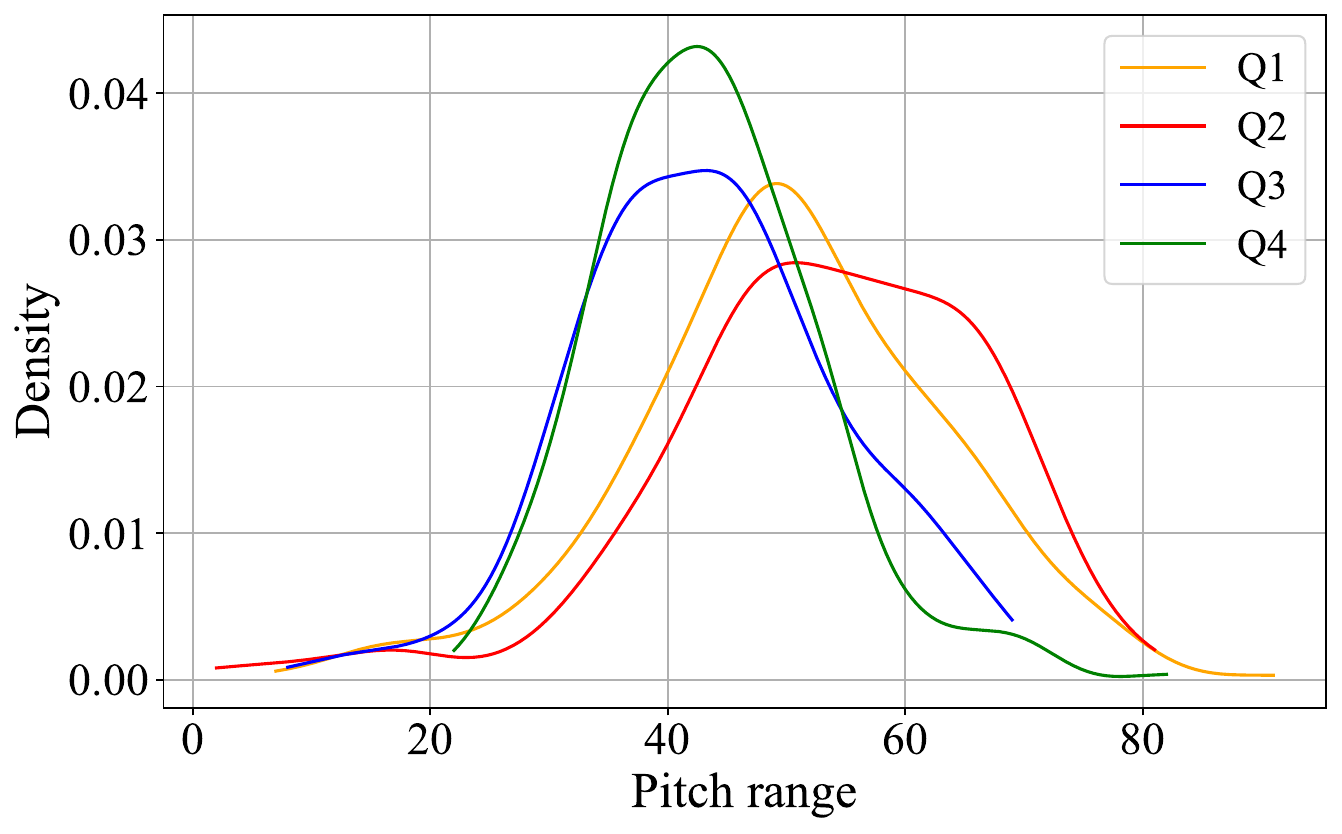}}\medskip
  \end{minipage}\hspace{3pt}
  \begin{minipage}[b]{0.24\linewidth}
    \centerline{\includegraphics[scale=0.2]{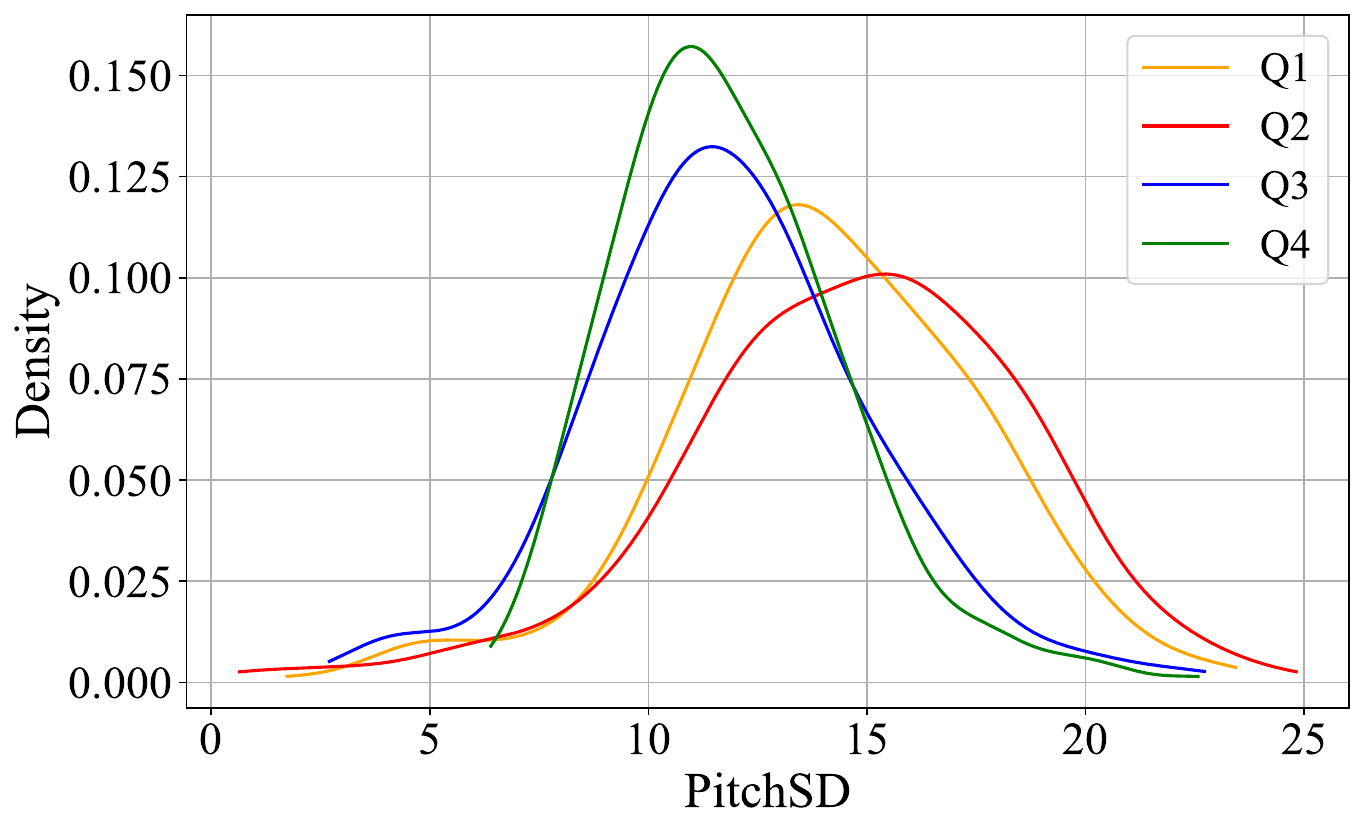}}\medskip
  \end{minipage}\hspace{5pt}
  \begin{minipage}[b]{0.24\linewidth}
    \centerline{\includegraphics[scale=0.2]{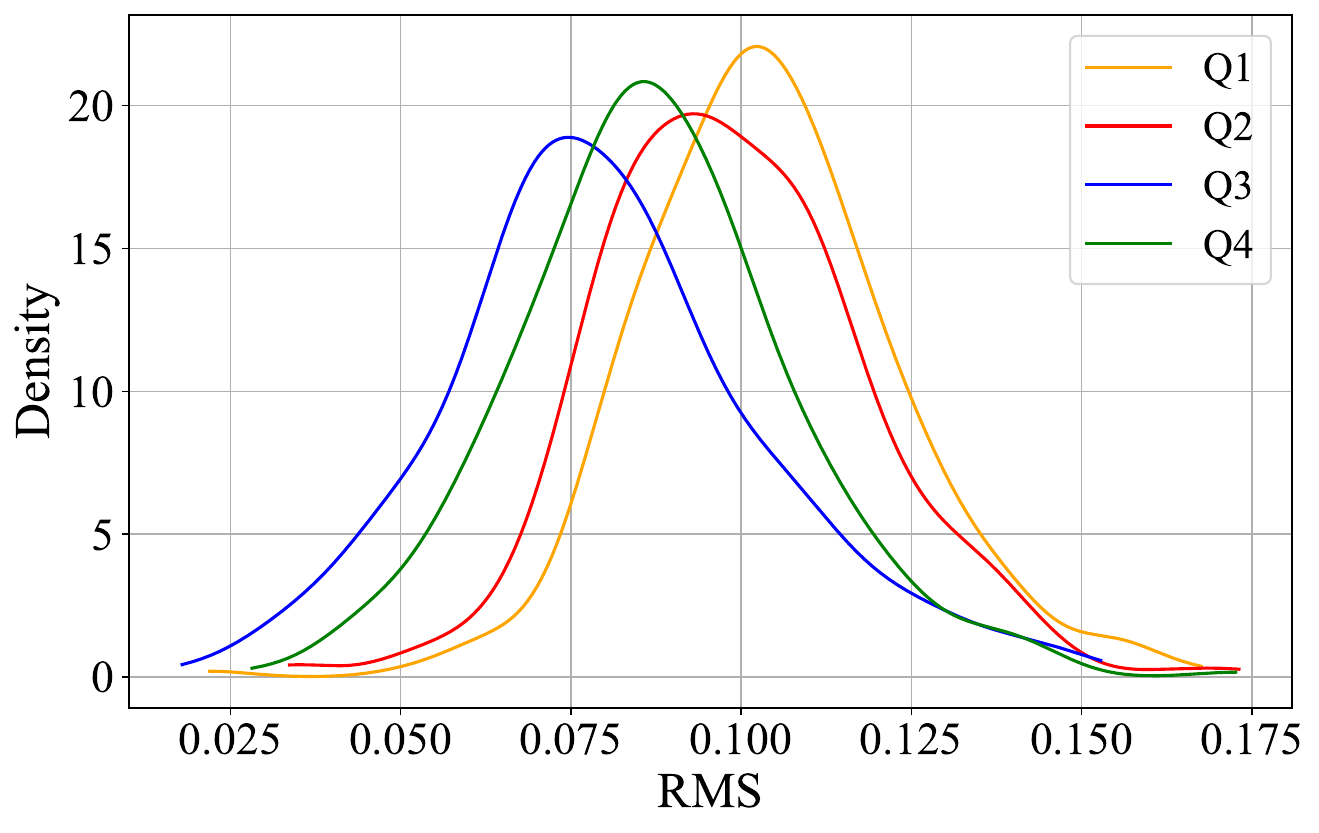}}\medskip
  \end{minipage}
  \begin{minipage}[b]{0.24\linewidth}
    \centerline{\includegraphics[scale=0.24]{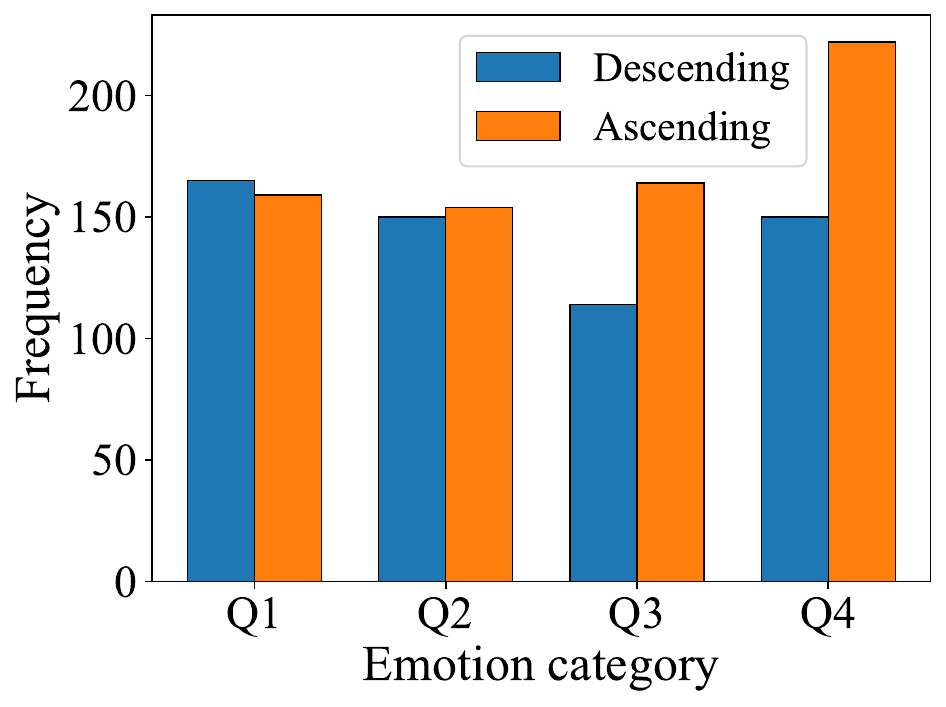}}\medskip
  \end{minipage}
  \caption{Gaussian KDE charts for Russell 4Q emotions over the six music-related features: key, tempo, average pitch, pitch range, pitchSD, and RMS, respectively (the six subplots on the left side); bar charts of Russell 4Q emotion frequency over modes and directions (the two subplots on the right side).}
  \label{fig:kdes}
\end{figure*}

We calculated the PCCs presented in Table \ref{tab:pearson} and plotted distribution charts in Fig. \ref{fig:kdes} between V/A and the eight features. These findings guided the subsequent data processing and experiment design for emotional control. The rough emotion labeling strategy of Rough4Q mentioned in Section \ref{subsec:4q} was also derived from them.

\subsection{Backbone Network} \label{subsec:model}
We selected pre-trained Tunesformer as the backbone network, which is designed specifically for generating ABC notations. Its input data format is divided into two parts: control code and ABC chars. The latter represents the conventional ABC notation music, with its data structure detailed in the document \textit{ABC Music Notation}\footnote{\href{https://trillian.mit.edu/~jc/music/abc/doc/ABC.html}{https://trillian.mit.edu/~jc/music/abc/doc/ABC.html}}. The former consists of four markers: S (number of sections), B (number of bars), E (edit distance similarity), and D (duration of the first section).

For the pre-training process, we consider a score dataset $S$ consisting of pairs $(x,y)$, where $x$ is the input musical score and $y$ is the target musical score. Each score is represented as a sequence of bar patches with each bar patch further decomposed into a sequence of characters $\{(b_1^1, b_2^1, \ldots, b_n^1), (b_1^2, b_2^2, \ldots, b_n^2) \ldots, (b_1^m, b_2^m, \ldots, b_n^m)\}$. The backbone is trained to predict each character token of the target score on the basis of the input score and the previously generated tokens in an autoregressive manner. Formally, the pre-training objective is to minimize the cross-entropy (CE) loss over all tokens in the target sequence:
\begin{equation}
  \mathcal{L}_{CE}(\theta)=-\sum_{(x, y)\in S}\sum_{i=1}^{m}\sum_{j=1}^{n}logp_{\theta}(b_j^i|x,b_{<j}^{<i})
  \label{eq:loss}
\end{equation}
where $p_\theta$ represents the probability distribution function of the backbone, parameterized by $\theta$, of predicting the correct character, and $b^i_j$ denotes the $j$-th character in the $i$-th bar patch of score $y$, $b^{<i}_{<j}$ refers to characters before the $j$-th character to the left of the $i$-th bar patch.

\begin{figure}[htbp]
  \centerline{
    \includegraphics[width=1.0\columnwidth]{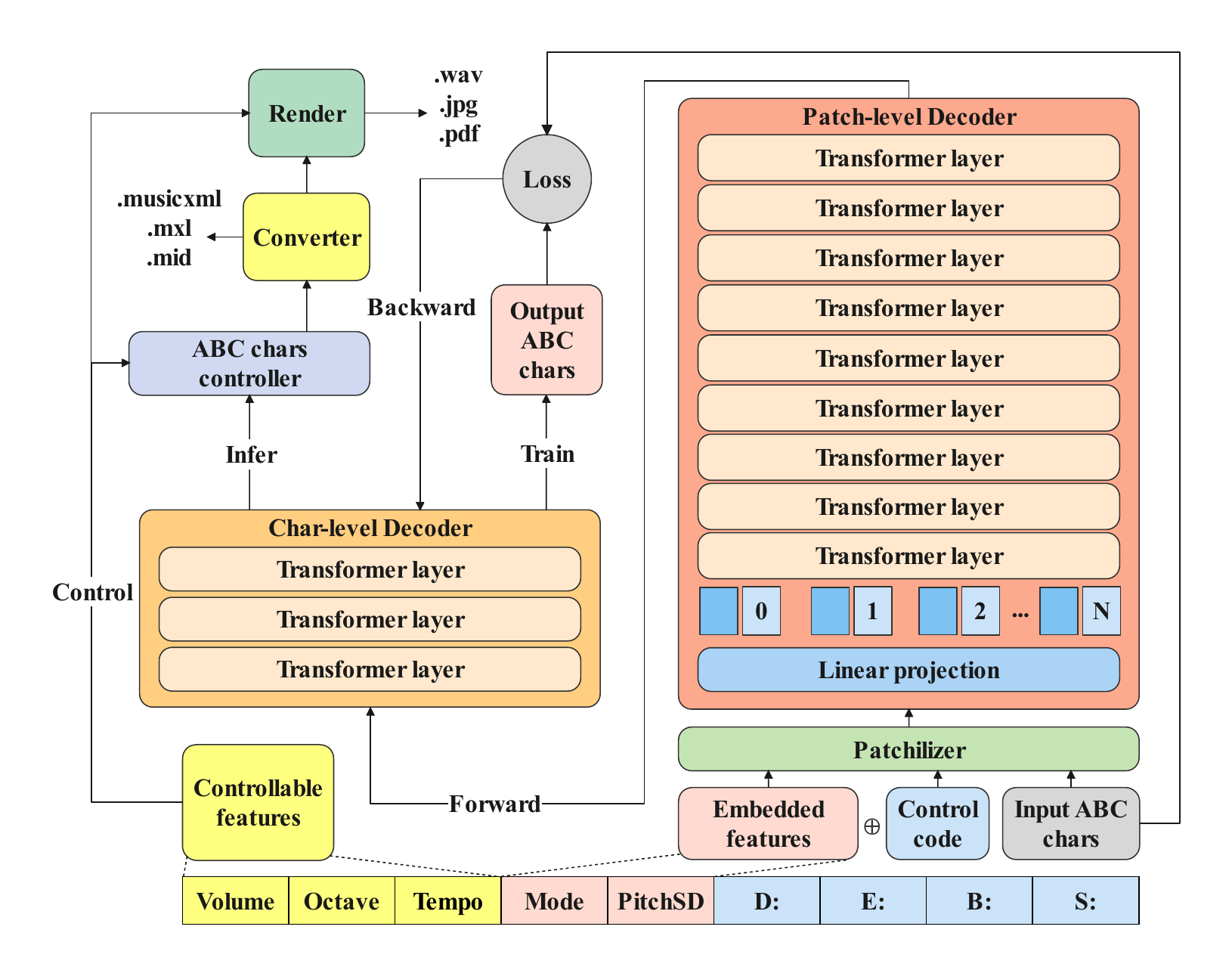}
  }
  \caption{The overall system architecture with training and inference branches of the backbone, whose below part outlines the musical features currently in use for control.}
  \label{fig:model}
\end{figure}

Building upon this backbone, we introduced the two embedded features and the three controllable features to achieve emotion-conditioned generation. The loss function used during fine-tuning is the same as that used during pre-training. The overall system architecture, integrating the backbone structure with the emotion control module, is illustrated in Fig. \ref{fig:model}.

\section{Experiments} \label{sec:exps}
\subsection{music21 Parsing Rate} \label{subsec:rsr}
The music21 parsing rate of a model refers to the proportion of musical scores generated by the model that can be successfully parsed by music21 without errors. This metric helps identify and filter out erroneous scores that may cause rendering failures and compromise the generation quality. We fine-tuned the backbone using the processed EMOPIA, processed VGMIDI, and Rough4Q datasets, which share the same data structure comprising three columns: control code, ABC chars, and 4Q label. Given that the attention mechanism of the transformer decoder is rightward, we parsed the 4Q label into string forms merged to the left of the control code.

We fine-tuned the backbone on these three datasets using a single H800 GPU in a Linux environment, with a batch size of 1. The training was early stopped once the evaluation loss during fine-tuning dropped below the minimum evaluation loss observed for the pre-trained model, and the model weights demonstrating the best performance were saved. This approach helps mitigate overfitting to prevent the model from generating melodies that are too similar to those in the training set.
We used the three fine-tuned models obtained above for inference, generating 100 pieces of ABC notation from each model. The music21 parsing rates of these scores were then calculated in Table \ref{tab:rate}. Owing to the inconsistent size of these three datasets, different sampling rates were used to control the amount of data used for fine-tuning for controlling the variables.
\begin{table}[htbp]
  \centering
  \footnotesize
  \caption{The comparison of music21 parsing rates among outputs from backbones fine-tuned by the processed EMOPIA, processed VGMIDI, and Rough4Q datasets.}
  \label{tab:rate}
  \begin{tabular}{|c|c|c|c|}
    \hline
    \textbf{Processed datasets} & \textbf{EMOPIA} & \textbf{VGMIDI} & \textbf{Rough4Q} \\ \hline
    Sampling rate (\%)          & 24.58           & 56.68           & 1.01             \\ \hline
    music21 parsing rate (\%)   & 28              & 75              & 99               \\ \hline
  \end{tabular}
\end{table}

The quality of a melody is a hardly defined metric that requires extensive subjective testing to minimize bias for obtaining reliable results. The music21 parsing rate, as a computable and objective measure, is one of the necessary conditions for quality. Therefore, this experiment can reflect the quality of the generated results to a certain extent.

\subsection{Ablation Study} \label{subsec:ablation}
On the basis of results in Table \ref{tab:rate}, models fine-tuned by processed EMOPIA and VGMIDI were found to have unsatisfactory music21 parsing rates for melody generation. Therefore, in subsequent experiments, we used a backbone fine-tuned with Rough4Q for further research. On the basis of statistical analysis in Table \ref{tab:pearson} and Fig. \ref{fig:kdes}, the guidance from music psychology conclusions, as well as our manual listening attempts, we selected the following five emotion control features, mode, tempo, pitchSD, volume (RMS), and octave (avg pitch), to design the emotion-conditioned generation template, where mode and pitchSD are managed through embedding, and the rest are controlled on the output stage of the backbone. The emotion control template for the five features is as follows:
\begin{itemize}
  \item \textbf{Q1}: major mode, high pitchSD, random tempo from 160-184 BPM (roughly around \textit{Allegro} -- \textit{Vivace}), no change in octave, volume increased by 5dB;
  \item \textbf{Q2}: minor mode, high pitchSD, random tempo from 184-228 BPM (roughly around \textit{Presto} -- \textit{Prestissimo}), pitches lowered by two octaves, volume increased by 10dB;
  \item \textbf{Q3}: minor mode, low pitchSD, random tempo from 40-69 BPM (roughly around \textit{Largo} -- \textit{Adagio}), pitches lowered by one octave, volume unchanged;
  \item \textbf{Q4}: major mode, low pitchSD, random tempo from 40-69 BPM, no change in octave and volume.
\end{itemize}

\begin{figure}[htbp]
  \begin{minipage}[b]{0.49\linewidth}
    \centering
    \centerline{\includegraphics[scale=0.3]{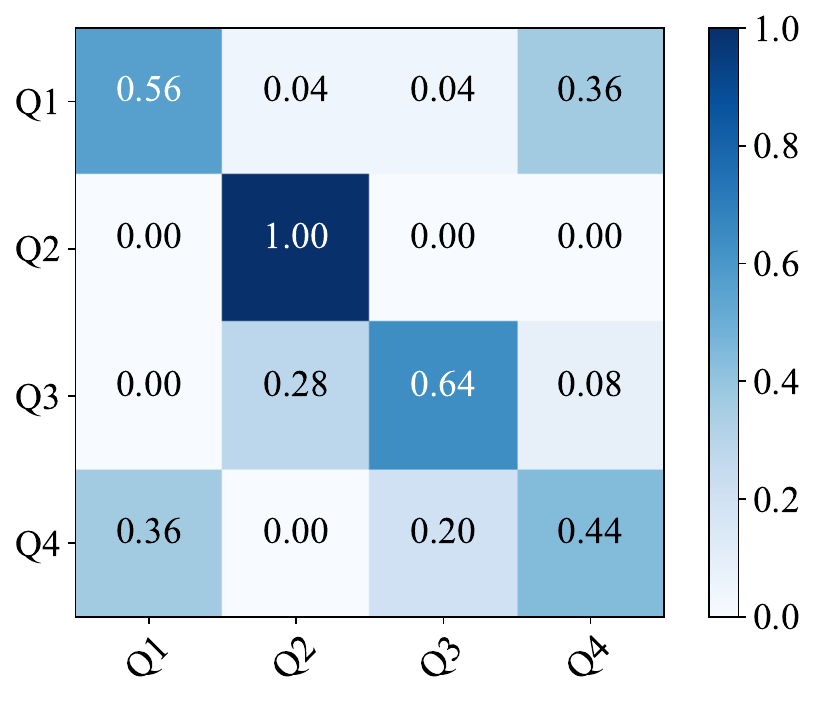}}
    \centerline{No tempo control}\medskip
  \end{minipage}
  \hfill
  \begin{minipage}[b]{0.49\linewidth}
    \centering
    \centerline{\includegraphics[scale=0.3]{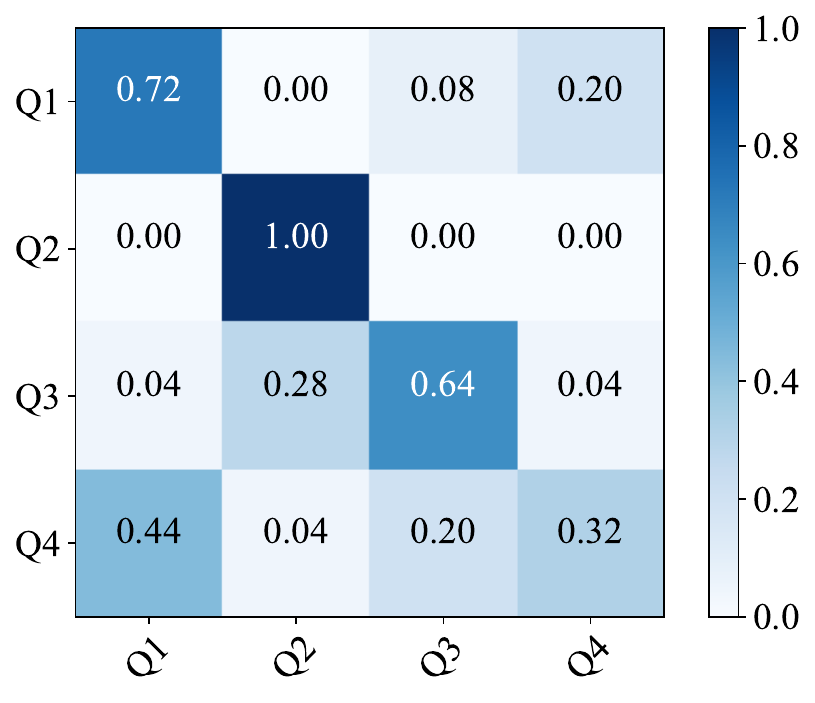}}
    \centerline{No pitchSD control}\medskip
  \end{minipage}

  \begin{minipage}[b]{0.49\linewidth}
    \centering
    \centerline{\includegraphics[scale=0.3]{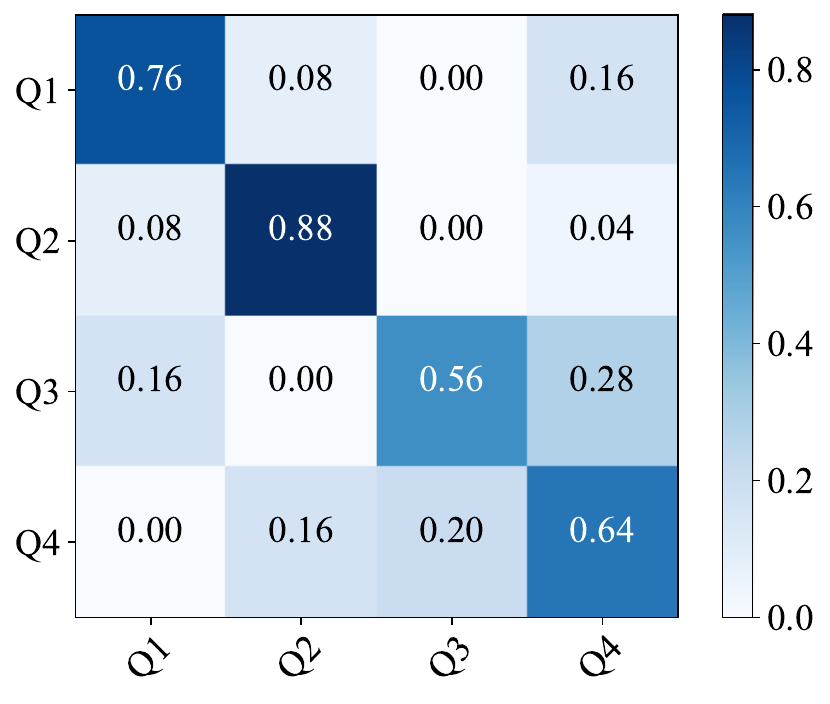}}
    \centerline{No mode control}\medskip
  \end{minipage}
  \hfill
  \begin{minipage}[b]{0.49\linewidth}
    \centering
    \centerline{\includegraphics[scale=0.3]{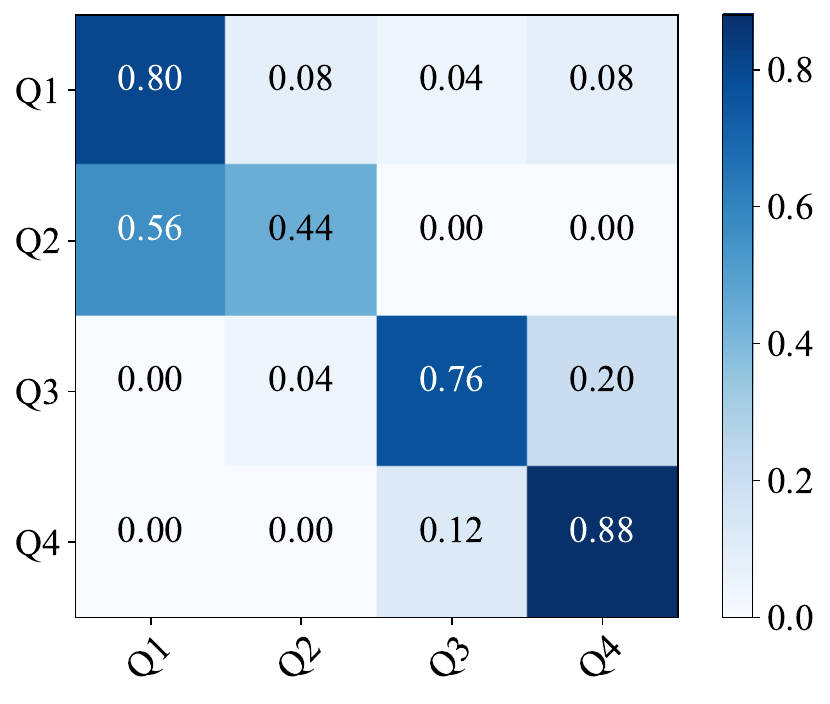}}
    \centerline{No octave control}\medskip
  \end{minipage}

  \begin{minipage}[b]{0.49\linewidth}
    \centering
    \centerline{\includegraphics[scale=0.3]{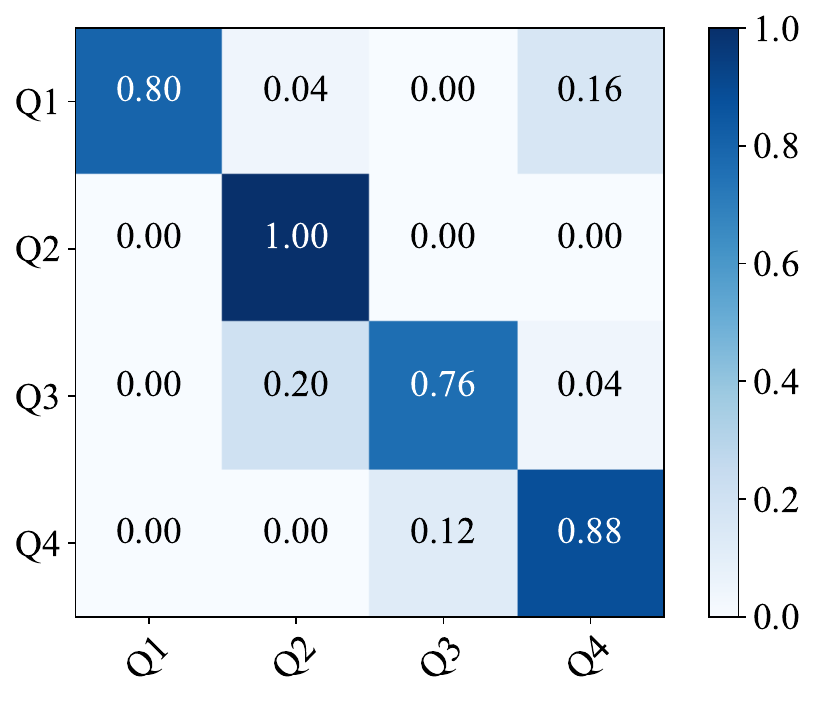}}
    \centerline{No volume control}\medskip
  \end{minipage}
  \hfill
  \begin{minipage}[b]{0.49\linewidth}
    \centering
    \centerline{\includegraphics[scale=0.3]{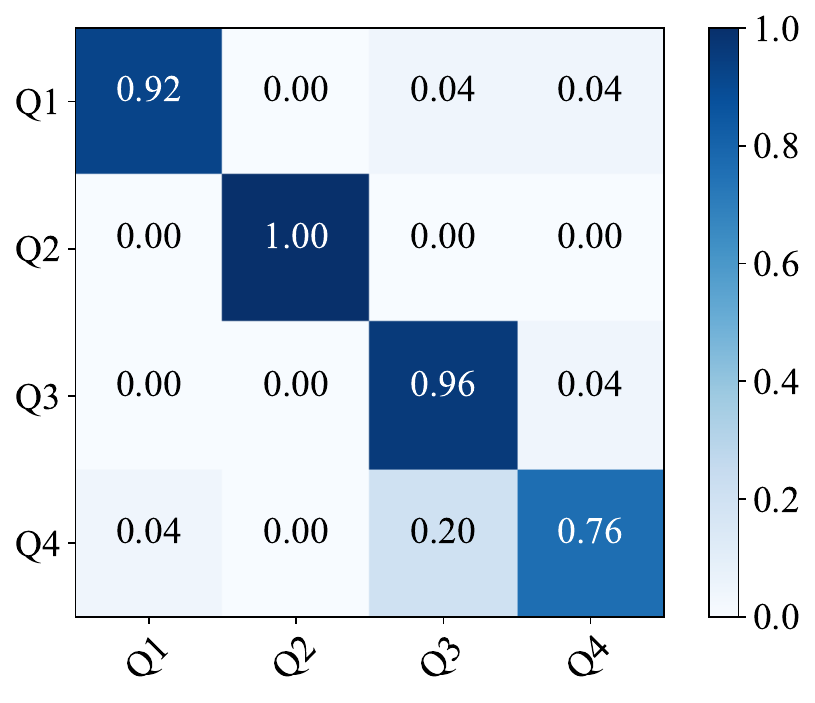}}
    \centerline{Full control}\medskip
  \end{minipage}
  \caption{Confusion matrices between human blind listening emotions of generated melodies and emotion prompts under full control and ablation options, where vertical axes represent the emotion prompts, and the horizontal axes represent the emotions labeled by the participants.}
  \label{fig:mats}
\end{figure}

Using this template, we generated 25 melodies for each emotion category, totaling 100 pieces. Under blind conditions, we first had three music enthusiasts listen to the pieces and label them according to their perceived 4Q emotions. To minimize subjective bias, we employed a two-of-three strategy: if at least two of the listeners identified a piece as a specific emotion, that emotion was considered the truth; if all three listeners provided three different responses, we randomized the discrepancies and replaced the three listeners to retake the test. We repeated that process until all discrepancies were resolved. After obtaining all labels from listeners, we compared them with the emotional condition in prompts to test the effectiveness of the emotional control template. The emotional generation accuracy of the current template was found to be 91\%. Additionally, we conducted ablation experiments on the five control conditions within the template in the same way, with all results in Table \ref{tab:exps} and Fig. \ref{fig:mats}. The ablation here refers to turning off the control of a specified feature in the template.
\begin{table}[htbp]
  \centering
  \footnotesize
  \caption{Performances of the generation model by comparing human blind listening emotions of generated melody and emotion prompts with ablation comparison.}
  \label{tab:exps}
  \begin{tabular}{ccccc}
    \hline
    \textbf{Ablation} & \textbf{Acc.(\%)} & \textbf{F1-score(\%)} & \textbf{Precision(\%)} & \textbf{Recall(\%)} \\\hline
    Tempo             & \textbf{66.0}     & 64.9                  & \textbf{64.8}          & \textbf{66.0}       \\
    PitchSD           & 67.0              & \textbf{64.8}         & 65.6                   & 67.0                \\
    Mode              & 71.0              & 70.8                  & 71.3                   & 71.0                \\
    Octave            & 72.0              & 71.2                  & 74.0                   & 72.0                \\
    Volume            & 86.0              & 85.9                  & 87.1                   & 86.0                \\\hline
    -                 & 91.0              & 90.9                  & 91.6                   & 91.0                \\\hline
  \end{tabular}
\end{table}

\section{Conclusions} \label{sec:conclusion}
From prior experiment results, directly converting EMOPIA and VGMIDI datasets into ABC notation for fine-tuning Tunesformer cannot ensure high music21 parsing rates on melody generation. In contrast, fine-tuning it on Rough4Q with controls by the musical feature template can be a more reliable method for emotion-conditioned melody generation. Key features such as mode, tempo, pitchSD, and RMS broadly align with music psychology findings, while avg pitch does not fully match them for the complexity of emotional judgments. The blind listening test validated the effectiveness of our template on the control of the aforementioned five features, showing that our method, not as a purely end-to-end emotional embedding approach, can still achieve around 91\% emotional generation performances. Ablation studies further validated that the control of the five features in the template can significantly affect generation performances, in which tempo, pitchSD, and mode play essential roles in particular.

\section*{Acknowledgment}
This work was supported by the National Social Science Fund of China (21ZD19), and the Special Program of the National Natural Science Foundation of China (T2341003).

\bibliographystyle{IEEEbib}
\bibliography{icme2025references}
\end{document}